\documentclass[12pt]{article}

\usepackage{graphicx}
\usepackage{amsmath}

\title{Imaging through scattering media by exploiting the optical memory effect: a tutorial}

\author{H. Penketh, and J. Bertolotti\footnote{j.bertolotti@exeter.ac.uk}}

\date{Department of Physics and Astronomy, University of Exeter, Exeter, Devon EX4 4QL, UK}

\begin{document}
\maketitle

\begin{abstract}
Scattering, especially multiple scattering, is a well known problem in imaging, ranging from astronomy to medicine. In particular it is often desirable to be able to perform non-invasive imaging through turbid and/or opaque media. Many different approaches have been proposed and tested through the years, each with their own advantages, disadvantages, and specific situations in which they work. In this tutorial we will show how knowledge of the correlations arising from the multiple scattering of light allows for non-invasive imaging through a strongly scattering layer, with particular attention on the practicalities of how to make such an experiment work.
\end{abstract}

%\maketitle

\section{Introduction}

Most objects around us are opaque. Which is good, because otherwise we wouldn't be able to see them. But the perk of being visible comes with the disadvantage of occluding anything behind them. There are many possible ways to deal with this problem, first and foremost is the option of physically removing the opaque obstacle, thus gaining direct line of sight to what we actually want to see, but they all come with their own advantages and disadvantages (for recent reviews of the topic see~\cite{reviewJBOK, ChoiReview2020, Roadmap2022}). In particular, techniques like optical clarification~\cite{clarificationreview} are invasive, meaning that they strongly modify the sample. This can be undesirable, so there is a push to develop non-invasive imaging techniques able to see things through an opaque medium.\\
In this paper we will focus on one such technique, which exploits the optical memory effect (which we will discuss in detail later) to reconstruct the image of an object hidden behind a scattering screen. This technique was originally developed under the name \textit{Stellar speckle interferometry}, in the context of ground-based astronomy, where the Earth's atmosphere acts as the scattering medium~\cite{Labeyrie1970}, but the same idea was later successfully applied to x-ray scattering~\cite{miao99, Abbey2011} and optical microscopy~\cite{Bertolotti2012, Katz2014}. This tutorial aims at providing a self-contained introduction to the technique, and a practical guide on how to set up an optical experiment and analyze the resulting data.

\section{Light scattering}
There are two optical phenomena that make objects visible to us: absorption and scattering. Green glass is mostly transparent (you can see through it), but since the other colors are largely absorbed by it, we perceive it as green. On the other hand clouds absorb very little, but they scatter sunlight, thus appearing white and opaque (dark clouds are still white, just thick enough that most sunlight is scattered back into space instead of reaching us.) As a rule of thumb absorption will make a signal weaker, while scattering will scramble it, with the scrambling being a much harder problem to tackle than the weakening (at least at the conceptual level). Following the time-honored tradition of breaking a difficult problem into smaller easier problems and tackling them one at the time, in the following we will forget about absorption and focus on the scrambling due to scattering.\\

\subsection{Modeling multiple scattering}
The theory of wave multiple scattering is now well established, with several reviews~\cite{RotterGiganReview, Cao2022, bartreviewlocalization} and textbooks~\cite{pingsheng, akkermansbook, carminatibook} devoted to it. In the following we will need only a few results from it, so here we will be content with a simplified intuitive picture. In a uniform and isotropic scattering medium, the average intensity of the unscattered light will decay exponentially with the distance $x$ from the source: $I_{\text{ballistic}}=I_0 e^{-x/\ell_s}$(Lambert-Beer law), where $\ell_s$ is the scattering mean free path, i.e. the typical distance between two scattering events. We can think of the scattering events to form an extended source, whose intensity will also decay exponentially, forming a source for twice-scattered light, etc. After $n$ scattering events, the total average intensity will be a exponential distribution convolved $n$ times with itself. This is not an easy calculation to do for an arbitrary $n$, but if we make the assumption that $n\gg 1$, i.e. we are in the multiple scattering regime, we can invoke the central limit theorem. This tells us that the $n$th convolution of identical distributions with finite variance will always converge to a Gaussian distribution.
Using the fact that in three dimensions the variance of the exponential distribution is $\sigma^2= 6 \ell_s^2$ we can write
\begin{equation}
    I(\mathbf r) = \frac{I (\mathbf{r}_0) }{\left( 2 \pi n 6\ell_s^2 \right)^{3/2}} e^{-\frac{|\mathbf r- \mathbf{r}_0|^2}{2 n 6\ell_s^2}}
    = \frac{I (\mathbf{r}_0) }{\left( 12 \pi \ell_s v t \right)^{3/2}} e^{-\frac{|\mathbf r- \mathbf{r}_0|^2}{12 \ell_s v t}} 
    = \frac{I (\mathbf{r}_0) }{\left( 4 \pi D t \right)^{3/2}} e^{-\frac{|\mathbf r- \mathbf{r}_0|^2}{4 D t}} \; ,
\end{equation}
where we estimated the number of scattering events as the total path $v t$ (where $v$ is the speed of light in the medium, and $t$ the elapsed time) divided by the scattering mean free path $\ell_s$, and defined the diffusion constant as $D=v \ell_s /3$. This is the well known bulk solution of the diffusion equation, which tells us that the propagation of light intensity in the multiple scattering regime is not very different from heat propagation. Which in retrospect is not overly surprising, as both can be thought as the average over Brownian random walks, and thus both satisfy the diffusion equation $\frac{\partial}{\partial t} I(\mathbf{r},t) = D \nabla^2 I(\mathbf{r}, t)$.\\
A full discussion of the properties of the diffusion equation is beyond the scope of this tutorial, but there are a couple of important features that are worth mentioning:
\begin{enumerate}
    \item The total amount of light transmitted through a scattering layer decreases with the layer thickness $L$ as $\ell_s/L$.
    \item A point source on one side of a scattering slab will produce a bell-shaped intensity distribution of width $\sim L$ on the other side. This explains why a thick enough scattering medium appears opaque.
\end{enumerate}
The first point is good news, as it tells us that some signal can pass through even relatively thick scattering media. On the other hand the second point is very bad news, as it makes difficult to form any good image through scattering media thicker than a few mm.

\begin{figure}[tb]
 \centering
  \includegraphics[width=0.8 \textwidth]{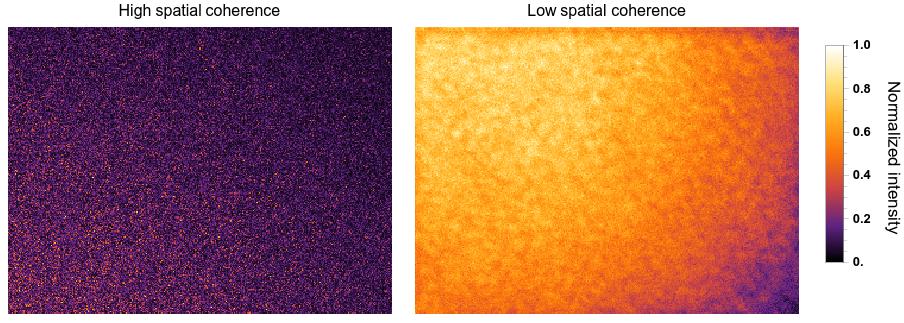}
   \caption{Left panel: When a coherent beam of light passes through a scattering medium, it forms a speckle pattern, composed by bright dots (speckles) surrounded by darker regions. Right panel: If the source spatial coherence is low (but its temporal coherence is still high) the speckle pattern will not average out, but will take a shape that depends on the shape of the source.}
 \label{fig:speckle}
\end{figure}
\subsection{Interference and speckle}
Thankfully, not everything is lost, as in modeling light propagation in scattering media we forgot an important point: light is a wave, and thus it interferes. This doesn't seem like a big deal at first sight, but it has a number of important consequences. The first one is that the light transmitted through a scattering slab won't be a shapeless blob, but will have a lot of internal structure. To see why, we can think about the field propagation as entering the scattering medium, performing a random walk, and then exiting the medium. Light doesn't actually do that (thinking about photons performing a random walk is a common misconception~\cite{LambAntiPhoton1995}), but it turns out that the average over all possible random walks yields the same result as the proper calculation. Since the amount of phase accumulated depends on the path length, the average field at each point on the output surface will be the sum of a lot of random terms. 
% It would be nice to break this big paragraph somewhere near here if possible, as what follows is quite a lot to take in
To perform this sum we can invoke the central limit theorem (again) and find that both the real and imaginary parts of the field are normally distributed. By making a change of variable from real and imaginary part, to intensity and phase, we find that the phase is uniformly distributed, and the intensity follows the exponential distribution $p(I)= \frac{1}{\left\langle I \right\rangle}e^{-I / \left\langle I \right\rangle}$ (this is for polarized light. If the light is not polarized, the result is slightly more complicated~\cite{goodman}), where $\left\langle . \right\rangle$ represents the average.
An important property of $p(I)$ is that its standard deviation is equal to its average, meaning that the intensity pattern after the scattering medium will fluctuate spatially from a maximum to zero, taking the shape of bright patches (speckles) surrounded by dark patches (see fig.\ref{fig:speckle}). This speckle pattern appears to be random, but actually encodes a lot of information about both the scattering medium and the light illuminating it.

\subsection{The optical memory effect}
\begin{figure}[tb]
 \centering
  \includegraphics[width=0.6 \textwidth]{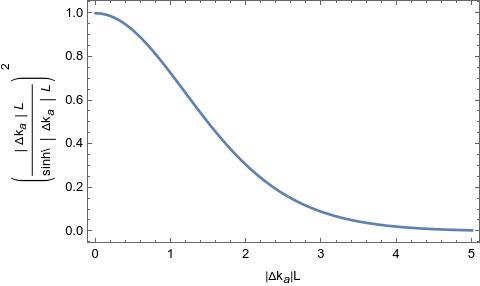}
   \caption{The speckle correlation due to the optical memory effect decreases rapidly if $|\Delta k_a| \, L \gg 1$, but for small values of $|\Delta k_a| \, L$ there is a perfect correlation.}
 \label{fig:memoryeffect}
\end{figure}
Despite how it appears, speckle patterns actually encode a lot of information about both the scattering medium and the light illuminating it. Teasing out some of this information will be the goal of the rest of this tutorial paper, and to do so we need to investigate speckle correlation. The fact that speckle is correlated means that the intensity at one point is not independent from the intensity at another point, as in the simplified model we used above. So we need a slightly better model.\\
We assume our scattering medium is a slab of infinite dimensions in both $x$ and $y$, and with a thickness $L$ along $z$. We then define the correlation between the intensity due to an incident plane wave with wavevector $\mathbf{k}_a$, emerging in the direction $\mathbf{k}_b$ ($I_{a,b}$) and the intensity due to an incident plane wave with wavevector $\mathbf{k}_{a^{\prime}}$, emerging in the direction $\mathbf{k}_{b^{\prime}}$ ($I_{a^{\prime},b^{\prime}}$) as
\begin{equation}
\label{eq:CI}
    \mathcal{C}^I_{a,a^{\prime},b,b^{\prime}} = \frac{\left\langle \delta I_{a,b}\, \delta I_{a^{\prime},b^{\prime}} \right\rangle}{\left\langle I_{a,b} \right\rangle \left\langle I_{a^{\prime},b^{\prime}} \right\rangle} \; ,
\end{equation}
where $\delta I = I- \left\langle I \right\rangle$ represent the fluctuations from the average. The calculation is not difficult but it is long, so we refer the interested reader to the literature~\cite{Feng1988, Freund1990} and simply state the result that, if we are far away from the Anderson localization transition, the leading term has the form
\begin{equation}\label{eq:c1}
    \mathcal{C}^{(1)}_{a,a^{\prime},b,b^{\prime}} =  \delta_{\Delta k_a , \Delta k_b} \; \left( \frac{|\Delta k_a| \, L}{\sinh |\Delta k_a| L} \right)^2 \, ,
\end{equation}
where $\Delta k_a = \mathbf{k}_a - \mathbf{k}_{a^{\prime}}$, and  $\delta$ is a Kronecker delta. This formula can be interpreted as follows: if we illuminate a scattering slab with a plane wave of wavevector $\mathbf{k}_a$, we produce a speckle pattern in transmission (and one in reflection). As shown in Fig.~\ref{fig:memoryeffect}, if now we change the angle of incidence by a small amount $|\Delta k_a|$, so that $\left( \frac{|\Delta k_a| \, L}{\sinh |\Delta k_a| L} \right)^2 \sim 1$, the resulting speckle pattern will still be highly correlated with the previous one (i.e. they will look the same), but also rotated by $|\Delta k_b|=|\Delta k_a|$. This is known as the \textit{optical memory effect}. In the other limit, if $|\Delta k_a| \, L \gg 1$, the correlation will be small, and the two speckle patterns will be very different from each other. As a rule of thumb, the angular range over which this effect is significant, is of the order of $\lambda /L$. If we illuminate with a finite-sized beam instead of a plane wave, the Kronecker delta in equation \ref{eq:c1} must be replaced with the square of the Fourier transform of the incident beam profile, which will determine the speckle shape and extension~\cite{akkermansbook}. It is worth noticing that if we look in reflection instead of in transmission, we get the same correlation term, but instead of $L$ we need to use $\ell_s$ in equation \ref{eq:c1}, which leads to a much bigger effective range for the memory effect.\\
This is by far not the only correlation that arises from wave scattering through a slab, but it is the strongest (albeit short-ranged), and it will enable us to perform non-invasive imaging through a scattering medium.

\section{Speckle interferometry theory}
Consider an object that either emits or reflects light with an intensity described by the function $O(\mathbf{r}_0)$, hidden behind a scattering screen of thickness $L$. If the light coming from the object is broadband (i.e. its temporal coherence is too low) the speckle patterns generated by each frequency will average out, but if the light has a long coherence length (or, equivalently, if our detection is narrow-band enough), the light coming from the point $\mathbf{r}_0$ on the object will produce a speckle pattern $S(\mathbf{r}_0, \mathbf{r}_d)$ at position $\mathbf{r}_d$~\cite{goodmanSpeckle}. If the spatial coherence of the source $O$ is low enough,
% could we provide a half a sentence definition of spatial coherence, or should that be assumed known? %%it would take a full page to explain it. Will put there a reference
these speckle patterns will not interfere, but their intensities will sum, resulting in a measured intensity
\begin{equation}
\label{eq:measuredI}
    I(\mathbf{r}_d) = \int O(\mathbf{r}_{o}) \, S(\mathbf{r}_{o},\mathbf{r}_{d})\, d^2\mathbf{r}_{o} ,
\end{equation}
i.e. the speckle pattern $S$ acts as a point spread function. This intensity $I(\mathbf{r}_d)$ can be measured, but doesn't really resemble the shape of the object we are interested in imaging (see fig.\ref{fig:speckle}). To proceed we take an autocorrelation of the measured intensity:
\begin{equation}
\label{eq:intensityautocorrelation}
    \begin{aligned}
        &[I \star I](\boldsymbol{\Delta r}_{d}) = \int I(\mathbf{r}_{d}) \, I(\mathbf{r}_{d}+\boldsymbol{\Delta r}_{d}) d^2\mathbf{r}_{d} =\\       &=
        \int \left[ \left( \int O(\mathbf{r}_{o}) \, S(\mathbf{r}_{o},\mathbf{r}_{d}) d^2\mathbf{r}_{o} \right) \left( \int O(\mathbf{y}_{o}) \, S(\mathbf{y}_{o},\mathbf{r}_{d}+\boldsymbol{\Delta r}_{d}) d^2\mathbf{y}_{o} \right) \right]\, d^2\mathbf{r}_{d} =\\
        &= \iint O(\mathbf{r}_{o}) O(\mathbf{r}_{o}) \left( \int S(\mathbf{r}_{o},\mathbf{r}_{d})\, S(\mathbf{y}_{o},\mathbf{r}_{d}+\boldsymbol{\Delta r}_{d})\, d^2\mathbf{r}_{d}  \right) \, d^2\mathbf{r}_{o}  d^2\mathbf{y}_{o} \\
=& \int O(\mathbf{r}_{o})\, O(\mathbf{y}_{o})
\Big( \left[ S \star S \right] (\mathbf{r}_{o}, \mathbf{y}_{o},\boldsymbol{\Delta r}_{d}) \Big) d^2\mathbf{r}_{o}  d^2\mathbf{y}_{o} \, .
    \end{aligned}
\end{equation}
where $\mathbf{y}_0$ is a dummy variable and $\star$ represents the correlation product (notice that all functions here are real-valued, so we omit any complex conjugate). We would now like to rewrite $S \star S$ in terms of the speckle correlation $\mathcal{C}^{I}$. This can be done in 2 steps: first we need to specialize eq.\ref{eq:CI} to our particular case, where the intensities to be correlated are the speckle patterns generated by two different points ($\mathbf{r}_{o}$ and $\mathbf{y}_{o}$ ) and measured at two different points ($\mathbf{r}_{d}$ and $\mathbf{r}_{d}+\boldsymbol{\Delta r}_{d}$):
\begin{equation}
    \left\langle \delta I_{a,b}\, \delta I_{a^{\prime},b^{\prime}} \right\rangle \rightarrow \left\langle \delta S(\mathbf{r}_{o},\mathbf{r}_{d})\, \delta S(\mathbf{y}_{o},\mathbf{r}_{d}+\boldsymbol{\Delta r}_{d}) \right\rangle \; .
\end{equation}
% should it be obvious why we're now using the fluctuations delta I / delta S %%We are rewriting eq2, that has the deltas
Assuming that spatial and ensemble averaging are equivalent we can then rewrite the correlation function as
\begin{equation}
\label{eq:CvsS}
    \mathcal{C}^{I} (\mathbf{r}_o, \mathbf{y}_{o},\boldsymbol{\Delta r}_{d}) = \frac{\left[ \delta S \star \delta S \right](\mathbf{r}_o, \mathbf{y}_{o},\boldsymbol{\Delta r}_{d})}{\left\langle S \right\rangle^2} \simeq \mathcal{C}^{(1)} \, .
\end{equation}
The second step is to write $S \star S$ in terms of $\delta S \star \delta S$:
\begin{equation}
\label{eq:autocorrelationdeltaS}
    \begin{aligned}
    &\delta S \star \delta S = \left( S- \left\langle S \right\rangle \right) \star \left( S- \left\langle S \right\rangle \right) =
    S \star S + \left\langle S \right\rangle \star\left\langle S \right\rangle  - 2 S\star \left\langle S \right\rangle=\\
    &= \int S(\mathbf{r}_{o},\mathbf{r}_{d})\, S(\mathbf{y}_{o},\mathbf{r}_{d}+\boldsymbol{\Delta r}_{d})\, \text{d}^2\mathbf{r}_{d} +
    \int \left\langle S \right\rangle^2\, \text{d}^2\mathbf{r}_{d} 
    -2 \int S(\mathbf{r}_{o},\mathbf{r}_{d})\, \left\langle S \right\rangle\, \text{d}^2\mathbf{r}_{d}=\\
    &= S\star S - \left\langle S \right\rangle^2 A \quad \Rightarrow S\star S = \delta S \star \delta S + \left\langle S \right\rangle^2 A \; ,
    \end{aligned}
\end{equation}
where $\left\langle S \right\rangle =\frac{\int S(\mathbf{r}_{o},\mathbf{r}_{d}) \, \text{d}^2\mathbf{r}_{d}}{A}$ is the average speckle intensity, and $A=\int \text{d}^2\mathbf{r}_{d}$ is the area covered by the speckle pattern.\\

We can now put together eq.\ref{eq:intensityautocorrelation}, eq.\ref{eq:CvsS}, and eq.\ref{eq:autocorrelationdeltaS} to obtain
\begin{equation}
    \begin{aligned}
        &[I \star I](\boldsymbol{\Delta r}_{d}) = \int O(\mathbf{r}_{o})\, O(\mathbf{y}_{o})
\Big( \left[ \delta S \star \delta S \right] (\mathbf{r}_{o}, \mathbf{y}_{o},\boldsymbol{\Delta r}_{d}) + \left\langle S \right\rangle^2 A \Big) d^2\mathbf{r}_{o}  d^2\mathbf{y}_{o} =\\
&= \left\langle S \right\rangle^2 \int O(\mathbf{r}_{o})\, O(\mathbf{y}_{o})  \mathcal{C}^{I} (\mathbf{r}_{o}, \mathbf{y}_{o},\boldsymbol{\Delta r}_{d}) d^2\mathbf{r}_{o}  d^2\mathbf{y}_{o} + A \left\langle S \right\rangle^2 \int O(\mathbf{r}_{o})\, O(\mathbf{y}_{o}) d^2\mathbf{r}_{o}  d^2\mathbf{y}_{o} \; .
    \end{aligned}
\end{equation}
This is easier to read if we make the change of variables $\mathbf{y}_{o}=\mathbf{r}_{o}+ \boldsymbol{\Delta r}_{o}$ and we identify the first set of integrals as correlations and convolution products (which we will label with $\star$ and $\otimes$ respectively), and call $\left\| O \right\|^2 = \int O(\mathbf{r}_{o})\, O(\mathbf{y}_{o}) \text{d}^2\mathbf{r}_{o}  \text{d}^2\mathbf{y}_{o}$:
\begin{equation}
\label{eq:compactautocorrelation}
    \begin{aligned}
        &[I \star I](\boldsymbol{\Delta r}_{d}) = \left\langle S \right\rangle^2 \left( [O\star O] \otimes \mathcal{C}^I + A \left\| O \right\|^2 \right) \; .
    \end{aligned}
\end{equation}
Apart from a prefactor $\left\langle S \right\rangle^2$, which only depends on how much light is illuminating the object $O$, eq.\ref{eq:compactautocorrelation} tells us that, by taking the autocorrelation of the intensity image $I$ we measure, we obtain the autocorrelation of the unknown object (convolved with the correlation $\mathcal{C}^{I}$), plus a constant background. This is general for any speckle correlation $\mathcal{C}^{I}$, but not all correlations are equally useful to retrieve the shape of the object $O$. As discussed above, we want to exploit the optical memory effect, so we can approximate
\begin{equation}
\label{eq:rewrittenC}
    \mathcal{C}^{I} \simeq \mathcal{C}^{(1)}=  \delta_{\Delta k_a , \Delta k_b} \; \left( \frac{|\Delta k_a| \, L}{\sinh |\Delta k_a| L} \right)^2 =  \delta_{\boldsymbol{\Delta r}_{o} , \boldsymbol{\Delta r}_{d}} \; \left( \frac{\frac{2\pi}{\lambda}\frac{|\boldsymbol{\Delta r}_{o}|}{d} \, L}{\sinh \frac{2\pi}{\lambda}\frac{|\boldsymbol{\Delta r}_{o}|}{d} L} \right)^2
\end{equation}
where $d$ is the distance between the object and the scattering layer and we made a small angle approximation (if the angles are big, the second factor in $\mathcal{C}^{(1)}$ goes to zero), so $k_a \sim \frac{2\pi}{\lambda}\frac{\boldsymbol{\Delta r}_{o}}{d}$, with $\lambda$ being the wavelength of the incident beam.

\subsection{Phase retrieval and the Gerchberg–Saxton algorithm}
\label{section:phaseretrieval}
Even in the most ideal case, where the object sits comfortably within the optical memory effect range and our illumination beam is wide enough that we can approximate it as a plane wave, the best we can directly extract from the intensity we measure is the autocorrelation of the object, not the shape of the object itself. In a few cases this is enough, e.g. binary stars seen from a ground-based telescope are usually too blurred by the atmosphere inhomogeneities to be able to resolve their angular distance, but if one measures in a sufficiently narrow band and for a sufficiently short time, one gets a speckly image whose autocorrelation is the autocorrelation of two small dots~\cite{Dainty1975}. And the autocorrelation of two small dots is 3 small dots, where the distance from the side ones to the central one is exactly the same as the distance between the two stars. While this explains why astronomers developed this approach, in many cases the objects to be imaged are too complex for any useful information to be gained by just looking at the autocorrelation of the object.\\

What one would like to do is to invert the autocorrelation and extract $O$. The problem here is that autocorrelation is a lossy operation, so it is not an invertible operator. Thankfully, if we can make assumptions about $O$ (e.g. in our case $O$ is real-valued and positive) there are well established techniques to find an approximation to $O$ starting from $O\star O$. Even better, if $O$ is at least two-dimensional (and since here we are dealing with images, we satisfy this criterion), the solution found by these techniques is unique~\cite{Fienup:78}.
We are not going to prove why these techniques work here, but we will discuss how they work and how to implement them.\\
Thanks to the Wiener–Khinchin theorem we know that the Fourier transform of the autocorrelation of a function is equal to the modulus squared of the Fourier transform of the function itself, i.e. $F[O\star O] = |F[O]|^2$, and thus we can directly extract the modulus of the Fourier transform from the autocorrelation. What we still need to be able to reconstruct $O$ is the phase of the Fourier transform, but almost all possible phases will result in a image that violates our initial assumptions when we perform the inverse Fourier transform, with the only exception being the \textit{correct} phase.\\

To iteratively search for this phase, the Gerchberg–Saxton algorithm starts from a guess of the true object $O$ (the better the guess the faster the convergence, but even very bad guesses will eventually converge to the correct solution), Fourier transform it, substitute the (wrong) amplitude with the one extracted from the autocorrelation, and Fourier transform back. Since the phase was wrong, what we get is not a real and positive function, but to nudge the algorithm in the right direction we can set to zero every pixel in the image that does not satisfy these constraints. We now have a slightly better guess than we had before, and we can repeat the whole process again and again, until the autocorrelation we measured and the autocorrelation of our new best guess are close enough to satisfy us. As one can imagine, setting to zero every pixel that doesn't satisfy our constraints is harsh, and while the algorithm works, it converges slowly and tends to get stuck for long periods of time. A better solution is to nudge slightly the values of the pixels that don't satisfy our constraints at each iteration, but of course there is an innumerable number of ways to do that, and choosing the best one is not easy. Thankfully, this hard work has already been done, and it is now generally accepted that the so-called \textit{hybrid input-output algorithm} is the best practical option. To explain the difference, in the Gerchberg–Saxton algorithm (also known as the \textit{error reduction algorithm} in this context), the guess $g$ at iteration $k+1$ is updated to the new guess $g^{\prime}$ at all points $x$ that satisfy the constraint $\gamma$ and set to zero otherwise, i.e.
\begin{equation}
    \label{eq:errorreduction}
    g_{k+1}(x)= \left\{
\begin{aligned}
    g'(x) \quad & x \in \gamma\\
    0 \quad & x\notin \gamma
\end{aligned}
    \right. \; ,
\end{equation}
while in the hybrid input-output algorithm
\begin{equation}
    \label{eq:HIO}
    g_{k+1}(x)= \left\{
\begin{aligned}
    g'(x) \quad & x \in \gamma\\
    g_k(x)-\beta g'(x) \quad & x\notin \gamma
\end{aligned}
    \right. \; ,
\end{equation}
where $\beta$ is a parameter that can be freely adjusted, to optimize the convergence. Empirically, the hybrid input-output algorithm is more erratic than the error reduction one for $\beta \gg 1$, but also tends to get in the vicinity of the desired solution much faster. On the contrary, the error reduction algorithm tends to stagnate for long periods, but once it is close to the desired solution, it will converge to it very reliably. There is no foolproof recipes on how to use these algorithms, but we have found that cycling some iterations of the hybrind input-output algorithm while gradually decreasing the value of $\beta$ tends to work in most cases. Like with most iterative methods, the more iterations one manages to perform, the better the final result will be (on average), but since these iterations are computationally expensive, one has to make a judgement call and decide when to stop. 
Thankfully, modern GPU acceleration allows one to perform a large number of Fast Fourier Transforms on large images relatively fast, thus making this computational approach significantly faster than it was when it was first developed.\\
One final important point to discuss is what this algorithm can \textit{not} do: autocorrelations do not contain any information about absolute position, just relative position between the various points composing the image, so this algorithm can never retrieve the absolute position of the imaged object. Furthermore, the autocorrelation of a real function is always centrosymmetric, so the algorithm can not distinguish between the image of the object and the same image flipped.

\section{A simple experimental implementation}
As this is a tutorial, we will focus on a minimal implementation of the experimental apparatus, which is both cheap and easy to build and run. The main components are:
\begin{description}
    \item[Light source] To satisfy the assumptions of eq.\ref{eq:measuredI} we need the temporal coherence of the signal to be large enough to generate a speckle pattern, but the spatial coherence to be low enough that the speckle patterns generated by different point will not interfere. This can be achieved in many ways. One is to have the object itself to be fluorescent, which automatically gives us low spatial coherence, and detect the signal through a narrow band filter to increase the temporal coherence~\cite{Bertolotti2012,Hofer2018}. Another, simpler, way is to start with a coherent light source (e.g. a laser), reduce its spatial coherence with e.g. a spinning diffuser, and use a mask as the object~\cite{Katz2014}. Depending on the specific arrangement, you might need some optics to direct the light in the desired direction.
    \item[Object] Depending on your light source you might need a fluorescent sample, or a mask that simply allows some of the light to pass through. An important, and often under-appreciated, point is that you want the whole object to fit within the optical memory effect range (which, if $L\gg\lambda$, can be very small), otherwise you will able to measure only a part of the autocorrelation $O\star O$ and the iterative algorithm will fail to reconstruct the object.
    \item[Scattering layer] To make the experiment simple you want all the incident light to be scattered, but at the same time the thickness to be small (so that the optical memory effect range is large). The simplest solution is to use a ground glass diffuser, which effectively acts as a random phase mask of negligible thickness.
    \item[Detection] We want to measure the speckle pattern in the far field, so position your camera such that each speckle spot is bigger than a pixel, but not much bigger. If you have constraints in where you can put the camera, some optics might be necessary. Since, as per eq.\ref{eq:compactautocorrelation}, we want to measure a relatively small signal sitting on a large background, an 8-bit camera is likely to not have enough dynamic range. We suggest to have at least a 12-bit dynamic range, but otherwise any camera where you can access the raw data (i.e. many smartphones are excluded) will suffice.
\end{description}

\subsection{A practical example}
\begin{figure}[tb]
 \centering
  \includegraphics[width=0.6 \textwidth]{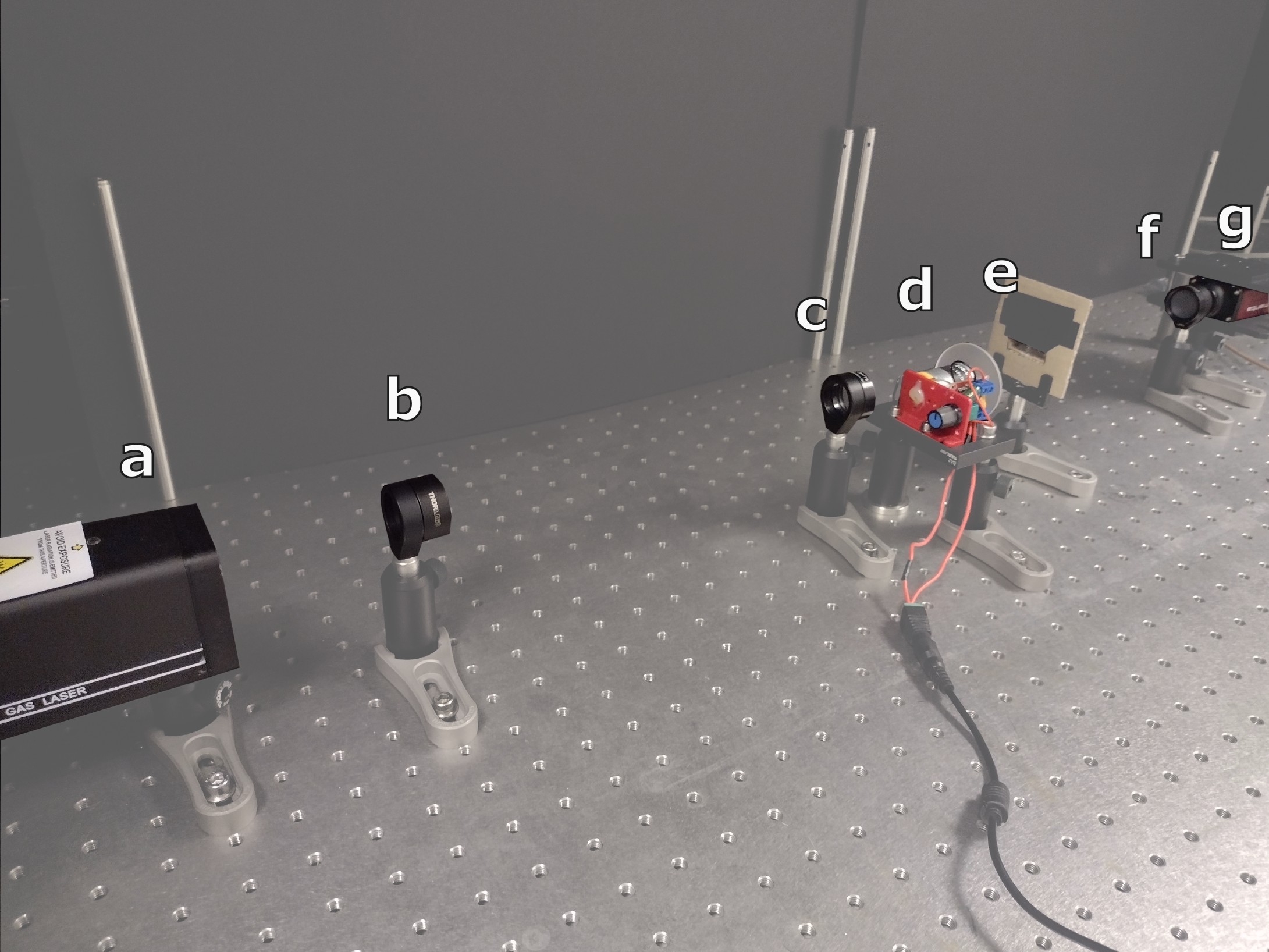}
   \caption{Photo of the experimental apparatus with the components highlighted.\\ a: Light source (1mW CW polarized HeNe laser).\\ b+c: Beam expander made from two lenses with focal length -25mm and 300mm respectively.\\ d: Home-made spinning diffuser to reduce the spatial coherence of the beam.\\ e: Sample (a piece of aluminum foil with 3 holes).\\ f: Scattering layer (220 grit ground glass optical diffuser).\\ g: Camera (Allied Vision Manta G-125B).}
 \label{fig:setup}
\end{figure}
In fig.\ref{fig:setup} we show a basic implementation of the experiment, that nevertheless contains all the features of more complex apparatuses. A laser beam with long coherence time is expanded and roughly collimated using two lenses to form a telescope. A spinning optical diffuser is used to reduce the spatial coherence of the laser source, and to effectively provide uniform illumination of the object when a single camera frame integrates over many realizations of the disorder. The diffuser increases the divergence of the of the expanded beam, hence perfecting the collimation of the beam expander is not crucial. Notice that the spinning diffuser is the only piece of this experiment that is not an \textit{off the shelf} component. There are a few commercially available speckle reducers. However, as we are not concerned with mechanical stability and vibrations here, the simple solution of a ground glass diffuser mounted on a small electric motor is sufficient.\\
The simplest possible choice for a sample is a screen that doesn't let any light pass through (e.g. a layer of aluminum foil) with a pattern carved into it. Three holes arranged in a slightly irregular triangle is the simplest non trivial pattern, and thus the one we use here.\\
The distance $d$ between the sample and the scattering layer is important because, as per eq.\ref{eq:rewrittenC}, it determines the angular size of the object as seen from the scattering layer, and thus whether it will fit in the memory effect range. Finally, a camera is placed after the scattering layer to collect the (scattered) light. This distance, combined with $d$, will determine how large the object will appear after the reconstruction~\cite{Hofer2018}.

\subsection{Data analysis}
\begin{figure}[tb]
 \centering
  \includegraphics[width=0.9 \textwidth]{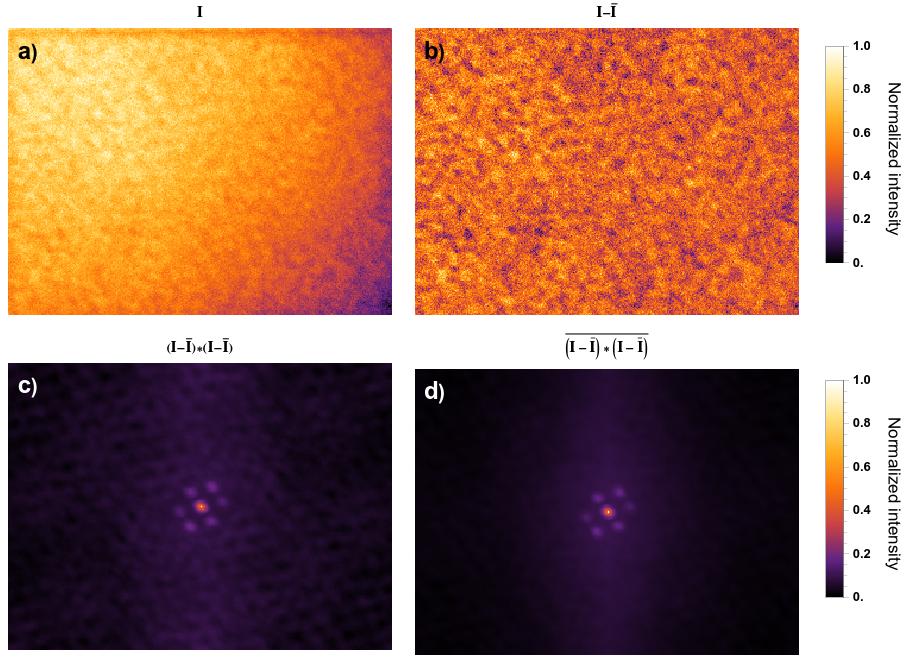}
   \caption{Data processing: a) Typical raw data measured by the apparatus with the nonuniform background clearly visible. b) Raw measurement minus the average measurements, which now sits on a flat background. c) Autocorrelation of the raw image with the background subtracted. The autocorrelation of the three dots is clearly visible. d) Average of the autocorrelations.}
 \label{fig:dataprocessing}
\end{figure}

In fig.\ref{fig:speckle} a typical raw image for both the case when the spinning diffuser is not moving, and when it is spinning are shown. In the first case we see a very fine-grained but high-contract speckle pattern, while in the second case, as per eq.\ref{eq:measuredI}, we see a relatively slow-varying but low-contrast image. This is the convolution of the object $O$ (in our case, three dots arranged in a triangle) and the unknown speckle $S$. This data needs to be cleaned before it can be used.\\
By looking at fig.\ref{fig:speckle} it is immediately visible that the raw data sits on an uneven background. This is due to an imperfect illumination and collection geometry, and while it can be ameliorated with a more careful alignment, it can never be completely removed. To correct for it we can take multiple measurements with different realizations of the scattering potential, and in this example we took 8 separate measurements of $I$. For the setup shown in fig.\ref{fig:setup} one can obtain it by rotating the diffuser (labelled as \textit{f} in the figure) by a few degrees each time. In more realistic experiments the scattering layer might be dynamic and change in time like a biological tissue, so one can simply take successive shots with the camera. Since the background is constant but the signal isn't, we can estimate the background by taking the average of our measurements $\bar{I}$, and subtract it from all the measurements (see fig.\ref{fig:dataprocessing}b).
We can now proceed to autocorrelate our measurements as in fig.\ref{fig:dataprocessing}c. As expected from eq.\ref{eq:compactautocorrelation} we can clearly see the autocorrelation of three dots arranged in a triangle, which is given by a central bright spot and six dots arranged around it in a hexagon. But contrary to what we expected from eq.\ref{eq:compactautocorrelation} we can see a lot of unwanted low-intensity structure in the background. This is due to the fact that we are only measuring a finite part of the signal, and thus the integral that gives rise to $S\star S$ in eq.\ref{eq:intensityautocorrelation} doesn't perform a perfect average. Since we have a number of autocorrelations $(I-\bar I)\star(I-\bar I)$, we can ameliorate this problem by averaging them to obtain $\overline{(I-\bar I)\star(I-\bar I)}$ (fig.\ref{fig:dataprocessing}d). Even after this processing the autocorrelation still suffer from a few problems, which need to be addressed before we can start the phase retrieval process to recover the object's shape. First of all, since the autocorrelation of white noise is a Dirac delta, and our measurements are unavoidably noisy, fig.\ref{fig:dataprocessing}d has a spike (a single pixel) in the very centre. As we know that this is just an artefact we can easily remove it by changing the value of that specific pixel to the value of a neighbour pixel. Another problem is that, since we subtracted the average measurement, now the autocorrelation can have negative values, which goes against our original assumption that we are essentially measuring the autocorrelation of a positive object $O$. The minimal correction we can make is to set every negative pixel to zero, or to add a constant background such that the autocorrelation is positive everywhere. This leads us to another, harder to solve problem: eq.\ref{eq:compactautocorrelation} tells us that what we are measuring is, even in the best case scenario, proportional to the autocorrelation of the unknown object plus a background, but we have no way to reliably estimate this background. Sometimes this is a major problem, and sometimes it isn't, depending on the data, but there is no real recipe on how to subtract that background beyond guesstimating it and try several values. In this example we will not subtract any further background.

\begin{figure}[tb]
 \centering
  \includegraphics[width=0.9 \textwidth]{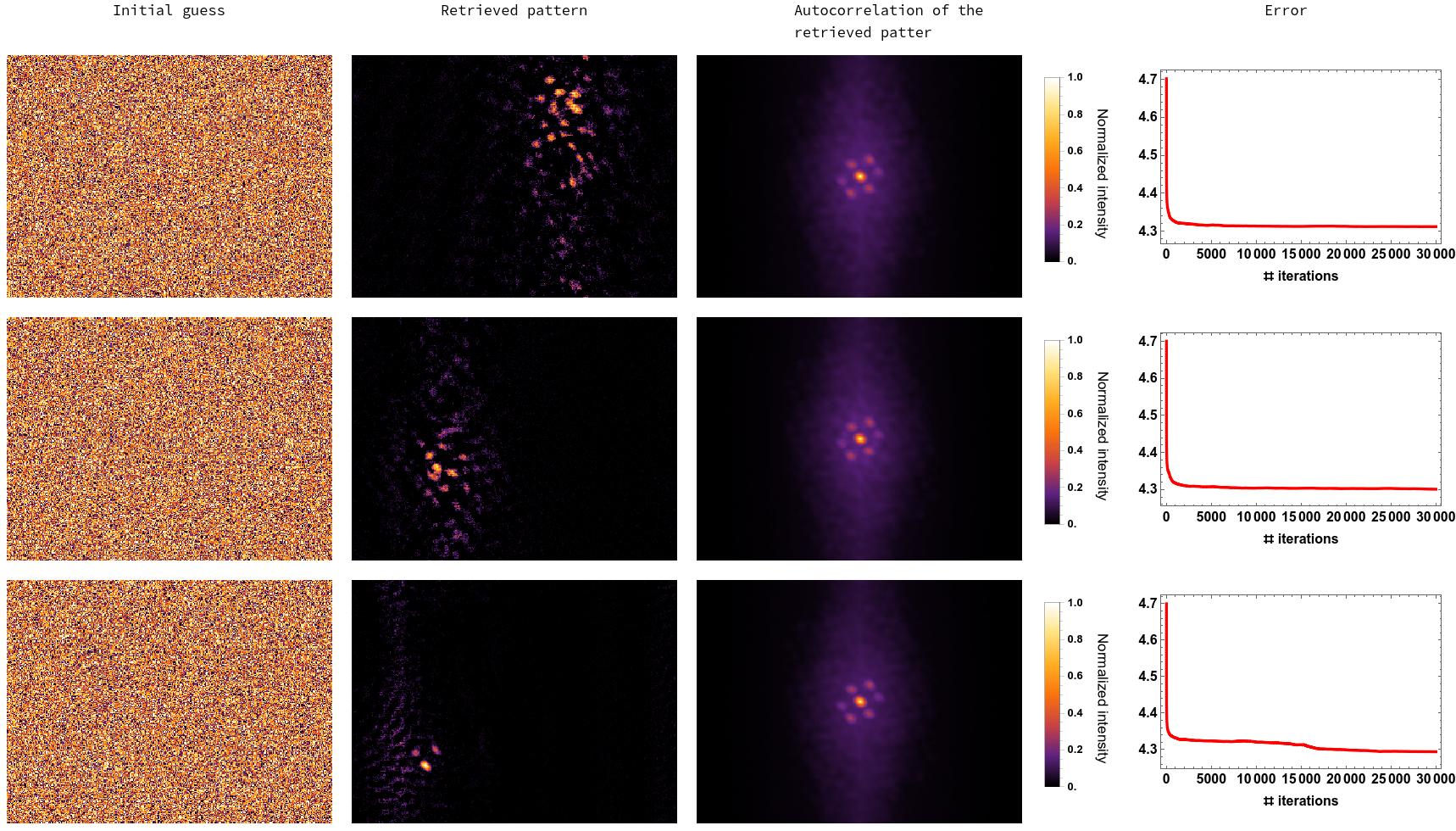}
   \caption{Phase retrieval using the Gerchberg–Saxton algorithm. Three runs of 30,000 iterations each, showing the initial random guess, the retrieved patter, the autocorrelation of the retrieved pattern, and the error as a function of the iteration number.}
 \label{fig:ER}
\end{figure}
\subsubsection{Practical phase retrieval}
In section~\ref{section:phaseretrieval} we introduced the Gerchberg–Saxton algorithm and its variation, the hybrid input-output algorithm. Implementing either of them as computer code is not exceptionally difficult, but there are a few details that is worth discussing.\\
Due to the finite precision of numerical evaluations, it is very likely that the inverse Fourier transform will produce a small imaginary part even when it should be zero, and if we implement naively the constraint that $O$ must be real, no element in the matrix representing the image will satisfy this constraint. One could impose a (arbitrary) threshold on the imaginary part to check if a given pixel satisfy the constraint or not, but in most cases it is easier to just set the imaginary part to zero and only select the pixel that do not satisfy the positivity of $O$.\\
Fig.\ref{fig:ER} shows three different runs of 30,000 iterations of the Gerchberg–Saxton algorithm defined in eq.\ref{eq:errorreduction} starting from 3 different initial guesses. In this case we assumed we had no information about the object, and used random numbers uniformly distributed between 0 and 1 as the guesses; a better initial guess will result in a faster convergence. We can track how the retrieval is proceeding by comparing the autocorrelation of the image at the $k$th iteration with the measured autocorrelation and define an error function estimate as
\begin{equation}
\label{eq:error}
\text{error}=\frac{\sum_{\text{pixels}} \left\lvert g_k\star g_k - \overline{(I-\bar I)\star(I-\bar I)} \right\rvert}{\text{number of pixels}}  \; .
\end{equation}
By looking at how this error function decreases with the iterations we can get an idea about how much progress we are making. From fig.\ref{fig:ER} we can see that Gerchberg–Saxton algorithm progresses very quickly in the first few iterations, but then the improvement becomes negligible, and only one of the three random initial guesses resulted in the recovery of the 3 dots after 30,000 iterations. By looking carefully at the error function we can see that the main difference between the successful one and the unsuccessful ones is that the successful one, after no appreciable improvement for roughly 15,000 iterations, had another relatively fast (albeit small) decrease, before flattening out again. These long stretches of no progress are a common feature of this algorithm, and are due to the fact that there are many very different false solutions that produce almost identical autocorrelations, and the algorithm can take a long time to jump from one to a better one. A practical trick is to run the algorithm multiple times with different initial guesses for a reasonable amount of iterations each time, and than choose the run that resulted in the smallest error, instead of a single very long run.
Another important feature clearly visible in fig.\ref{fig:ER} is that, since the autocorrelation of a real and positive function is always centrosymmetric, the algorithm is not able to distinguish between $O$ and its mirror image. Similarly, an autocorrelation doesn't contain information about absolute position, so the position of the reconstructed image is essentially random.\\
\begin{figure}[tb]
 \centering
  \includegraphics[width=0.9 \textwidth]{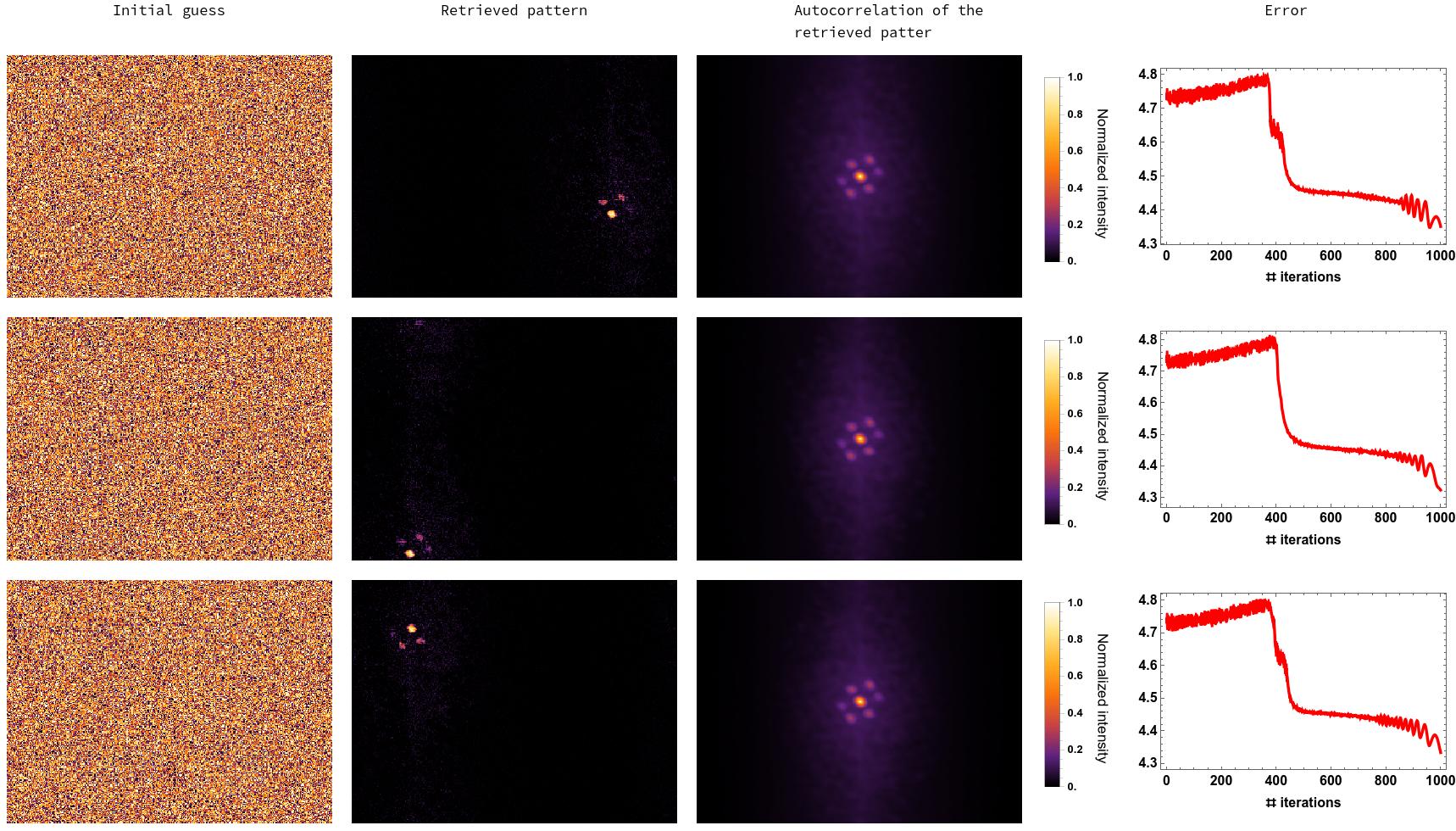}
   \caption{Phase retrieval using the hybrid input-output algorithm while linearly decreasing $\beta$ from 2 to 0. Three runs of 1,000 iterations each, showing the initial random guess, the retrieved patter, the autocorrelation of the retrieved pattern, and the error as a function of the iteration number.}
 \label{fig:HIO}
\end{figure}
The hybrid input-output algorithm, defined in eq.\ref{eq:HIO}, is designed to reduce the convergence time. The parameter $\beta$ decides how much each iteration is nudged in the right direction, so a reasonable strategy is to start with a relatively large value of $\beta$, so that the solution space can be sampled faster, and then gradually reduce it to zero. Fig.\ref{fig:HIO} shows three different runs of 1,000 iterations of the hybrid input-output algorithm starting from 3 different initial guesses, and with $\beta$ being linearly decreased from 2 to 0 with the number of iterations. Despite being 30 times shorter than the runs with the Gerchberg–Saxton algorithm, all three runs correctly converged to the three dots pattern, and all show a clear transition around $400$ iterations (i.e. $\beta\sim1.2$).
There is no agreed best way to choose $\beta$ or to combine these algorithms, but what we discussed here is an approach that tends to work fine for most cases.

\section{Conclusions}
Non-invasive imaging in strongly scattering media is unlikely to be a solvable problem in its generality, but different approaches that work under different conditions and in different cases are constantly being developed. Here we discussed a technique based on the optical memory effect, that allows non invasive imaging through a scattering layer. The autocorrelation of the object can be measured with a resolution given by the speckle correlation ($\mathcal{C}^I$ acts as a point spread function in eq.\ref{eq:compactautocorrelation}), and this autocorrelation numerically inverted using a phase retrieval algorithm. The main advantages of this approach is its simplicity, and the fact that it works even for extremely strongly (and/or dynamic) scattering layers. It can also work in a reflection geometry~\cite{Katz2014} or to image a fluorescent medium with a microscope~\cite{Hofer2018}. The phase retrieval is computationally expensive, but since it largely amounts to Fourier transforms and operations on a matrix elements, it can be sped up significantly by exploiting modern GPUs.\\
The main limitations of this approach is that the angular range of the optical memory effect is usually very small (see fig.\ref{fig:memoryeffect}), and thus the objects to be imaged will need to be either very small or very far away from the scattering layer, or only the central part of the autocorrelation will be measured.\\

\bibliographystyle{ieeetr}

\begin{thebibliography}{10}

\bibitem{reviewJBOK}
J.~Bertolotti and O.~Katz, ``Imaging in complex media,'' {\em Nature Physics},
  vol.~18, p.~1008, 2022.

\bibitem{ChoiReview2020}
S.~Yoon, M.~Kim, M.~Jang, Y.~Choi, W.~Choi, S.~Kang, and W.~Choi, ``Deep
  optical imaging within complex scattering media,'' {\em Nature Reviews
  Physics}, vol.~2, p.~141, 2020.

\bibitem{Roadmap2022}
S.~Gigan, O.~Katz, H.~B. de~Aguiar, E.~R. Andresen, A.~Aubry, J.~Bertolotti,
  E.~Bossy, D.~Bouchet, J.~Brake, S.~Brasselet, Y.~Bromberg, H.~Cao,
  T.~Chaigne, Z.~Cheng, W.~Choi, T.~Čižmár, M.~Cui, V.~R. Curtis,
  H.~Defienne, M.~Hofer, R.~Horisaki, R.~Horstmeyer, N.~Ji, A.~K. LaViolette,
  J.~Mertz, C.~Moser, A.~P. Mosk, N.~C. Pégard, R.~Piestun, S.~Popoff, D.~B.
  Phillips, D.~Psaltis, B.~Rahmani, H.~Rigneault, S.~Rotter, L.~Tian, I.~M.
  Vellekoop, L.~Waller, L.~Wang, T.~Weber, S.~Xiao, C.~Xu, A.~Yamilov, C.~Yang,
  and H.~Yılmaz, ``Roadmap on wavefront shaping and deep imaging in complex
  media,'' {\em Journal of Physics: Photonics}, vol.~4, p.~042501, 2022.

\bibitem{clarificationreview}
D.~S. Richardson and J.~W. Lichtman, ``Clarifying tissue clearing,'' {\em
  Cell}, vol.~162, p.~246, 2015.

\bibitem{Labeyrie1970}
A.~Labeyrie {\em Astron. Astrophys.}, vol.~6, p.~85, 1970.

\bibitem{miao99}
J.~Miao, P.~Charalambous, J.~Kirz, and D.~E. Sayre {\em Nature}, vol.~400,
  p.~342, 1999.

\bibitem{Abbey2011}
B.~Abbey, L.~W. Whitehead, H.~M. Quiney, D.~J. Vine, G.~A. Cadenazzi, C.~A.
  Henderson, K.~A. Nugent, E.~Balaur, C.~T. Putkunz, A.~G. Peele, G.~J.
  Williams, and I.~McNulty, ``Lensless imaging using broadband x-ray sources,''
  {\em Nature Photonics}, vol.~5, p.~420, 2011.

\bibitem{Bertolotti2012}
J.~Bertolotti, E.~van Putten, C.~Blum, A.~Lagendijk, W.~Vos, and A.~Mosk,
  ``Non-invasive imaging through opaque scattering layers,'' {\em Nature},
  vol.~491, p.~232, 2012.

\bibitem{Katz2014}
O.~Katz, P.~Heidmann, M.~Fink, and S.~Gigan, ``Non-invasive single-shot imaging
  through scattering layers and around corners via speckle correlations,'' {\em
  Nature Phot}, vol.~8, p.~784, 2014.

\bibitem{RotterGiganReview}
S.~Rotter and S.~Gigan, ``Light fields in complex media: Mesoscopic scattering
  meets wave control,'' {\em Rev. Mod. Phys.}, vol.~89, p.~015005, 2017.

\bibitem{Cao2022}
H.~Cao, A.~P. Mosk, and S.~Rotter, ``Shaping the propagation of light in
  complex media,'' {\em Nature Physics}, vol.~18, p.~994, 2022.

\bibitem{bartreviewlocalization}
B.~van Tiggelen, ``Localization of waves,'' in {\em Wave Diffusion in Complex
  Media} (J.~P. Fouque, ed.), NATO Science, Kluver, 1998.

\bibitem{pingsheng}
P.~Sheng, {\em Introduction to Wave Scattering, Localization and Mesoscopic
  Phenomena}.
\newblock Springer, 2010.

\bibitem{akkermansbook}
E.~Akkermans and G.~Montambaux, {\em Mesoscopic Physics of Electrons and
  Photons}.
\newblock Cambridge University Press, 2007.

\bibitem{carminatibook}
R.~Carminati and J.~C. Schotland, {\em Principles of scattering and transport
  of light}.
\newblock Cambridge University Press, 2021.

\bibitem{LambAntiPhoton1995}
W.~E. Lamb, ``Anti-photon,'' {\em Applied Physics B}, vol.~60, p.~77, 1995.

\bibitem{goodman}
J.~W. Goodman, {\em Statistical Optics}.
\newblock John Wiley \& Sons, 1985.

\bibitem{Feng1988}
S.~Feng, C.~Kane, P.~A. Lee, and A.~D. Stone, ``Correlations and fluctuations
  of coherent wave transmission through disordered media,'' {\em Phys. Rev.
  Lett.}, vol.~61, pp.~834--837, Aug 1988.

\bibitem{Freund1990}
I.~Freund, ``Looking through walls and around corners,'' {\em Physica A},
  vol.~168, pp.~49--65, 1990.

\bibitem{goodmanSpeckle}
J.~W. Goodman, {\em Speckle Phenomena in Optics: Theory and Applications}.
\newblock SPIE, 2020.

\bibitem{Dainty1975}
J.~C. Dainty, {\em Stellar Speckle Interferometry}, p.~255–280.
\newblock Springer Berlin Heidelberg, 1975.

\bibitem{Fienup:78}
J.~R. Fienup, ``Reconstruction of an object from the modulus of its fourier
  transform,'' {\em Opt. Lett.}, vol.~3, no.~1, p.~27, 1978.

\bibitem{Hofer2018}
M.~Hofer, C.~Soeller, S.~Brasselet, and J.~Bertolotti, ``Wide field
  fluorescence epi-microscopy behind a scattering medium enabled by speckle
  correlations,'' {\em Opt. Expr.}, vol.~26, p.~9866, 2018.

\end{thebibliography}

\end{document}